\documentclass[twocolumn,showpacs,amsmath,amstex,amssymb,mathfonts,prl]{revtex4-1}
\usepackage{graphicx}

\usepackage{tipa}
\usepackage{bm}

\usepackage{amsmath}
\usepackage{amssymb}
\usepackage{amsthm}
\usepackage{amsfonts}

\begin{document}

\title{Nature of Quasielectrons and the Continuum of Neutral Bulk Excitations in Laughlin Quantum Hall Fluids}

\author{ Bo Yang and F.D.M. Haldane}
\affiliation{Department of Physics, Princeton University, Princeton, NJ 08544, USA}

\date{August 21, 2013}
\begin{abstract}
We construct model wavefunctions for a family of single-quasielectron states supported by the $\nu=1/3$ fractional quantum Hall (FQH) fluid. The charge $e^*$ = $e/3$ quasielectron state is identified as a composite of a charge-$2e^*$ quasiparticle and a $-e^*$ quasihole, orbiting around their common center of charge with relative angular momentum $n\hbar > 0$,  and corresponds precisely to the ``composite fermion'' construction based on a filled $n=0$ Landau level plus an extra  particle in level $n > 0$.  An effective three-body model  (one $2e^*$ quasiparticle and  two $-e^*$ quasiholes)
is introduced to capture the essential physics of the neutral bulk excitations.
\end{abstract}

\maketitle 

Elementary excitations of the fractional quantum Hall (FQH) fluids not only are the building blocks of the multi-component FQH states, they also form neutral  bound states with an energy gap that characterizes FQH incompressibility.   In the Abelian $\nu$ = $1/m$ Laughlin  FQH states\cite{laughlin},  the quasihole states correspond to  insertion of an extra flux quanta through the two-dimensional ``Hall surface'', and are well-understood. Quasielectron excitations require addition of both electrons and flux quanta, and are in general more complex (i.e. one cannot take quasielectrons as ``anti-particles" of quasiholes). 

A  model wavefunction for a single quasielectron was first proposed by Laughlin\cite{laughlin}, and was later improved by Jain\cite{jain1}.    In Jain's ``composite fermion'' (CF) picture, the FQH state is described as an integer quantum Hall (IQHE) state of  CFs (electron-vortex composites)  in an effective magnetic field, where they fill in what Jain  has called ``$\Lambda$-levels'' ($\Lambda$Ls) \cite{jainbook}. 
A model CF wavefunction for a lowest-Landau-level (LLL) FQH state is obtained by multiplying the corresponding  Slater determinant IQHE state by  an even power of the Vandermonde determinant, followed by projection into the LLL.    The IQHE state corresponding to the $n$ = 1  $\Lambda$L quasielectron state of Ref.\cite{jain1} corresponds to a filled LLL plus a single electron in the second ($n$ = 1) Landau level.
This description of quasielectron states has been reformulated in the formalisms of conformal field theory\cite{hansson,rodriguez} and that of Jack polynomials\cite{bernevig}: though  both these descriptions seem very different from that of Ref.\cite{jain1},  all three constructions turn out to produce identical quasielectron states.

Model quasielectron states where  the extra CF is placed  in a  \textit{higher}  ($n > 1$) $\Lambda$L were  recently used in the construction\cite{jain2} of ``CF excitons" consisting of a quasielectron-quasihole bound state\cite{jain2}, in an attempt to explain experimental inelastic light-scattering observations\cite{pinczuk} that were interpreted as the splitting of the neutral bulk excitation spectrum at  long wavelengths.  Placing the extra CF in different $\Lambda$Ls leads to different species of exciton states.
Most constructions of the quasielectron wavefunctions are implemented in the  spherical geometry\cite{haldanesphere}, where many-particle states are characterized by a total
angular-momentum quantum number $L$, and FQH ground states have $L= 0$.    Charged excitations have
$L \sim  O(N_e)$, where $N_e$ is the number of electrons, while neutral excitations such as excitons have $L \sim O(1)$, with $L \ge 2$. Ref.\cite{jain2} studied the  energies of the ``$n > 1$ exciton'' bands as a function of $L$ (which  gives their energy as a function of momentum in the large-$N_e$ limit), and found that the exciton energy became independent of $n$ in the small-$L$ limit.  However, as pointed out here,  after the CF projection into the LLL, these  exciton states are not linearly independent, and in the $L=2$ and $L=3$ sector their  microscopic wavefunctions become \textit {identical}. This suggests that a better understanding of the nature of these ``excitons", or quasielectron-quasihole pairs, is needed.

In this Letter,  we show there is a family of $e/3$ quasielectron states that can themselves be interpreted as a composite of a $2e/3$ quasiparticle with a $-e/3$ quasihole orbiting around it.    The closest allowed orbit corresponds to  placement of the extra CF in the $n=1$ $\Lambda$L, and  has a much lower energy than
$n > 1$ states,  which are quasi-degenerate in the case of the model ``short-range pseudopotential'' interactions\cite{haldanesphere} for which the Laughlin state is the exact ground state.  The model wavefunctions of the family of single quasielectron states and the resulting neutral excitations (a charge $2e/3$ quasielectron and two $-e/3$ quasiholes) can also be obtained by a universal scheme, described here, based on the Jack polynomial formalism, which naturally explains the counting and linear dependence of various $\Lambda$L quasielectron and exciton states in different angular momentum sectors, seen in numerical studies\cite{jain2}.

We first recall the single quasielectron state for the Laughlin state constructed in \cite{bernevig}, which is identical to the one constructed in the CF  picture\cite{jain1}. For convenience we look at the fermionic Laughlin state at filling factor $\nu=1/3$. In the language of the fermionic generalization\cite{regnault} of the  Jack polynomials, where the clustering properties of the many-body wavefunction can be explicitly defined\cite{bh}, the ``root configuration'' is 
\begin{eqnarray}\label{qpl1}
\textsubbar{1}\textsubbar{1}0\textsubring{0}01001001001\cdots\quad L=N_e/2 .
\end{eqnarray}
In the ``hierarchy picture''\cite{haldanesphere}, this configuration can be identified as a single  ``elementary droplet''  (11000) of the $\nu = 2/5$  state ``daughter state''
in the background of its $\nu$ = 1/3 parent state (with elementary droplet 100). 
On the sphere, this single quasielectron state is in the total angular momentum sector $L=N_e/2$, where $N_e$ is the number of electrons. Each binary number represents an orbital on the sphere, arranged sequentially with the leftmost orbital at the north pole and the rightmost orbital at the south pole. Each ``1'' represents an occupied orbital, ``0'' an empty orbital. The root configuration of the ground state at $\nu=1/3$ has the clustering property that there is one electron in every three consecutive orbitals. Any deviation from that clustering property in the root configuration indicates the presence of a charge $e/3$ quasiparticle or a charge $-e/3$ quasihole, respectively indicated by a
bar $\textsubbar{1}$ or an open circle $\textsubring{0}$ below the occupation number in (\ref{qpl1}). 

The many-body wavefunction $|\psi_{N_e/2}^{\text{qe}}\rangle$ is given by\cite{yangbo}
\begin{eqnarray}\label{qpexpansion}
|\psi_{N_e/2}^{\text{qe}}\rangle=\sum_i\alpha_i|\psi^{s}_i\rangle+\beta c^\dagger_0c^\dagger_1|\psi^{J}\rangle
\end{eqnarray}
where $|\psi^s_i\rangle$ is the set of all basis squeezed from the root configuration\cite{bh} in (\ref{qpl1}) with at most \textit{one} of the first two orbitals occupied; $|\psi^J\rangle$ is the state corresponding to fermionic Jack polynomial $J^{-2}_{000001001001001\cdots}$, with an ``admissible''  root configuration  obtained by annihilating the first two electrons (in orbitals labeled 0 and 1) of the root configuration of (\ref{qpl1}); the coefficients $\alpha_i$ and $\beta$ can be uniquely fixed\cite{yangbo}  by imposing the highest weight condition $L^+|\psi^{\text{qe}}_{N_e/2}\rangle=0$, where $L^+$ is the raising operator of the total $L_z$ on the sphere.
It is noteworthy that for a basis derived in this way from such a root configuration, application of the highest-weight condition leads to a \textit{highly-overdetermined} set of
linear equations, which nevertheless have a (unique) solution.

Physically, the excess charge of the quasielectron state is concentrated at the north pole, and the state relaxes to the ground state far away from the north pole, as manifested by the Jack polynomial in (\ref{qpexpansion}), which is a zero energy state of the short-range pseudopotential model Hamiltonian.  From the positions of  the underscores and circles in (\ref{qpl1}) one can infer that the $e/3$ quasielectron state is a composite of a $2e/3$ quasiparticle (we here reserve the term ``quasielectron"   for excitations
with opposite charge to the quasihole) at the north pole, with a $-e/3$ quasihole orbiting around it. The key proposal of this Letter is to generalize (\ref{qpl1}) into a family of root configurations
\begin{eqnarray}\label{qprest}
&&\textsubbar{1}\textsubbar{1}0\textsubring{0}01001001001\cdots\qquad L=N_e/2\nonumber\\
&&\textsubbar{1}\textsubbar{1}0010\textsubring{0}01001001\cdots\qquad L=N_e/2+1\nonumber\\
&&\textsubbar{1}\textsubbar{1}0010010\textsubring{0}01001\cdots\qquad L=N_e/2+2\nonumber\\
&&\textsubbar{1}\textsubbar{1}0010010010\textsubring{0}01\cdots\qquad L=N_e/2+3  \quad \textit{etc.}
\end{eqnarray}
Above,  we list the first four states of the family. The same scheme used to define states with root configuration (\ref{qpexpansion})  uniquely defines  model wavefunctions for each of the root configurations in (\ref{qprest}). We verified that the quasielectron state in the $L=N_e/2+ (n-1)$ sector with $ n \ge 1$  is \textit{identical} to the quasielectron state constructed in the CF picture (with the lowest-Landau-level projection implemented 
exactly, using symbolic computer algebra for $N_e \le 6$), when the extra CF is in the $n${th} $\Lambda$L (the zero-energy ground state is given by the filled $n$ = 0 $\Lambda$L). The root configurations give  insight into the nature of the CF construction using  higher $\Lambda$Ls: the orbit of the quasihole around the $2e/3$ quasiparticle increases with increasing $\Lambda$L, defining the effective size of the quasielectron. For a finite system of $N_e$ electrons, the total number of single quasielectron states is $N_e-1$. In the thermodynamic limit where $N_e\rightarrow\infty$, for any finite $n >0$, a net excess charge $e/3$ remains   near  the origin.

\begin{figure}
\includegraphics[width=7.5cm]{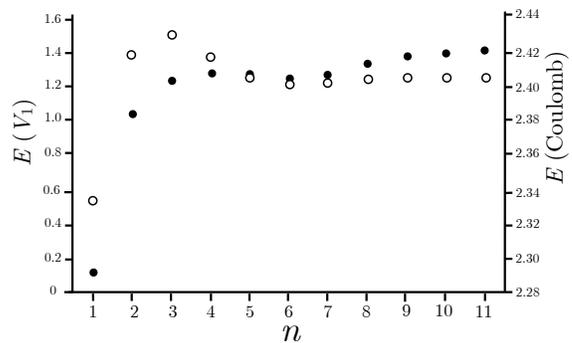}
\caption{Variational energies of the quasielectron states, evaluated with the $V_1$ pseudopotential Hamiltonian (open circles, left axis, units $V_1$) and the Coulomb interaction (solid circles, right axis, units  $e^2/4\pi \epsilon_0\epsilon l_B$). The system size is 12 electrons in 33 orbitals.  The index $n$ (the
``$\Lambda$-level'' index in the CF picture) is related to the  
 total angular momentum $L$ by $n-1$ = $L-N_e/2$, where $N_e$ is the number of particles.} 
\label{qe}
\end{figure}

We now evaluate the variational energies of this family of single quasielectron states. From Fig.(\ref{qe}), one can see that for both the short-range $V_1$ pseudopotential interaction and the more realistic Coulomb interaction, there is a large energy gap between the $L=N_e/2$ quasielectron and other quasielectron states in the family; the $L=N_e/2+ (n-1)$ states with $n>1$ are quasi-degenerate with similar energies, implying a small binding energy between the $2e/3$ quasiparticle and the $-e/3$ quasihole, once they are a few units of magnetic length away from each other. Thus from a dynamical point of view, for the Laughlin state there are effectively two basic excitations with the same sign of charge as the electron: (a) a $e/3$ quasielectron composite state made of a $2e/3$ quasiparticle and a tightly bound ($n$ = 1) quasihole; (b) a bare $2e/3$ quasiparticle.

 With the newly-defined family of quasielectrons, we now turn our attention to the neutral bulk excitations of the FQHE at $\nu=1/3$. Since the neutral excitations are made of quasielectron-quasihole pairs, different types of  single quasielectron states can lead to different types of single-pair neutral excitations. 
The root configurations of the neutral bulk excitations (see  table) are  obtained from those  in (\ref{qprest}) by inserting an additional empty orbital while  keeping intact the $2e/3$ quasiparticle (the $m=1$ pair $\textsubbar{1}\textsubbar{1}\ldots$ at the north pole). 
\begin{table}
\begin{tabular}{ll}
 $L=2$ & $L=3$\\
\textsubbar{1}\textsubbar{1}0\textsubring{0}\textsubring{0}01001001001001$\cdots\quad$&
\textsubbar{1}\textsubbar{1}0\textsubring{0}010\textsubring{0}01001001001001$\cdots$\\
 $L=4$ & $L=5$\\
\textsubbar{1}\textsubbar{1}0\textsubring{0}010010\textsubring{0}01001001$\cdots\quad$&
\textsubbar{1}\textsubbar{1}0\textsubring{0}010010010\textsubring{0}01001001$\cdots$\\
\textsubbar{1}\textsubbar{1}0010\textsubring{0}\textsubring{0}01001001001$\cdots\quad$&
\textsubbar{1}\textsubbar{1}0010\textsubring{0}010\textsubring{0}01001001001$\cdots$\\
$L=6$ & $ L=7$\\
\textsubbar{1}\textsubbar{1}0\textsubring{0}010010010010\textsubring{0}01$\cdots\quad$ &
\textsubbar{1}\textsubbar{1}0\textsubring{0}010010010010010\textsubring{0}01$\cdots$\\
\textsubbar{1}\textsubbar{1}0010\textsubring{0}010010\textsubring{0}01001$\cdots\quad$&  
\textsubbar{1}\textsubbar{1}0010\textsubring{0}010010010\textsubring{0}01001$\cdots$\\
\textsubbar{1}\textsubbar{1}0010010\textsubring{0}\textsubring{0}01001001$\cdots\quad$& 
 \textsubbar{1}\textsubbar{1}0010010\textsubring{0}010\textsubring  {0}01001001$\cdots$
\end{tabular}
\end{table}
The number of states in the $L=N$ sector is $[N/2]$, the greatest integer less or equal to $N/2$. The first root configuration in each $L$ sector  corresponds to the  magneto-roton mode\cite{gmp}. The many-body wavefunctions of the additional neutral modes can be constructed just like the quasielectron states: we found the resulting wavefunctions are equivalent to the ``CF  excitons"  of Ref.\cite{jain2}. Note for $L=2$ and $L=3$, each sector contains only \textit{one} state corresponding to the magneto-roton mode. Thus, in these sectors, apparently-different ``CF exciton" states  become identical after projection  (perhaps explaining why the Monte Carlo CF calculations of Ref.\cite{jain2} found their variational energies appeared to coincide in this limit).  

\begin{figure}
\includegraphics[width=8.5cm]{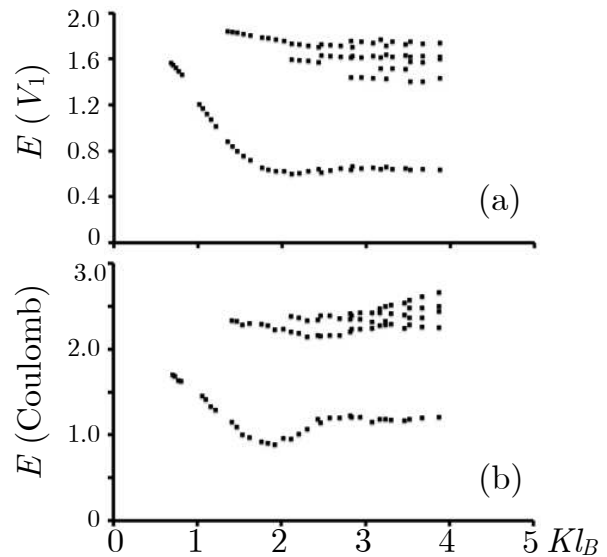}
\caption{The energy spectrum (as a function of momentum $K$) obtained (a) by diagonalizing the $V_1$ pseudopotential Hamiltonian  and (b) the Coulomb interaction (units $e^2/4\pi \epsilon_0\epsilon l_B$) within the Hilbert space of the model wavefunctions obtained from (\ref{nm}).The data is generated from system sizes ranging from 8 to 12 particles.}
\label{nm}
\end{figure}

We also diagonalized the two-body-interaction Hamiltonian within the subspace spanned by these neutral excitation modes, for both short-range ($V_1$) psedopotential\cite{haldanesphere} and Coulomb interactions. A clear separation of the magneto-roton mode and the continuum of other neutral excitations is seen in Fig.(\ref{nm}), though all the modes seem to merge together into the continuum in the long wavelength limit. We want to emphasize that states in this continuum consists of \textit{one} quasielectron-quasihole pair\cite{yangbo}, distinguishing them from the multi-roton continuum starting at an  energy double the roton-minimum gap\cite{gmp}, which involves \textit{two} quasielectrons.  These additional neutral excitations are buried in the multi-roton continuum, and in the thermodynamic limit the number of single-pair neutral excitations is also macroscopic.

To further understand the property of the single-pair neutral excitations for the Laughlin state, especially in the long wavelength limit, we first look at quasielectrons in (\ref{qprest}). Denoting the guiding center coordinate of the quasiparticle with charge  $2e^*$ by $R_0^a$ and that of the quasihole with charge $-e^*$ by $R_1^a$, where $a=x,y$ is the spatial index, we have the following commutation relations
\begin{eqnarray}\label{comm1}
[R^a_0,R^b_0]=-{\textstyle\frac{1}{2}}i\epsilon^{ab}l^{*2}_B,\quad [R_1^a,R_1^b]=i\epsilon^{ab}l^{*2}_B .
\end{eqnarray}
Here $\epsilon^{ab}$ is the $2D$ antisymmetrization symbol and $l^*_B=\sqrt{\hbar /|e^*B|}$ is the effective magnetic length. Assuming rotational invariance, the two-body interaction Hamiltonian is given by
\begin{eqnarray}\label{hamph}
\mathcal H_{ph}=\int \frac{d^2ql_B^2}{2\pi} U(|q|)e^{iq_a\left(R_0^a-R^a_1\right)},
\end{eqnarray}
where  the two-body interaction has a ``pseudopotential expansion'' in terms of Laguerre polynomials:
\begin{eqnarray}\label{hamphexpand}
U(|q|)={\textstyle\frac{1}{2}}\sum_nE_n L_n\left({\textstyle\frac{1}{4}}|ql^*_B|^2\right)e^{-\frac{1}{8}|ql^*_B|^2}.
\end{eqnarray}
From Fig.(\ref{qe}) all pseudopotentials are close to zero except for $E_1$. Without  loss of generality we now set $E_1$ = $-1$,  $E_{n>1}$ = 0.
 To project out the $n=0$ quasielectron state (as such a state  would correspond to the invalid root configuration $1\textsubbar{0}1001001001\cdots$, where the highest-weight charge $2e^*$ quasiparticle $\textsubbar{1}\textsubbar{1}$ has been destroyed), we also require $E_0=\infty$. 
This set of pseudopotentials $E_n$ in (\ref{hamph}) describes qualitatively the energetics of the quasielectrons in (\ref{qprest}).

Next, denoting the guiding center coordinates of the two quasiholes by $R_1^a,R_2^a$, the two-body interaction Hamiltonian for the quasiholes is given by
\begin{eqnarray}\label{hamhh}
\mathcal H_{hh}=\int \frac{d^2ql_B^2}{2\pi}V(|q|)e^{iq_a\left(R_1^a-R_2^a\right)} .
\end{eqnarray}
With rotational invariance one can also expand the  two-body interaction $V(|q|)$ in terms of pseudopotentials $V_n$
\begin{eqnarray}\label{hamhhexpand}
V(|q|)=2\sum_nV_n L_n(|ql^*_B|^2)e^{-\frac{1}{2}|ql^*_B|^2} .
\end{eqnarray}
Combining (\ref{hamph}) and (\ref{hamhh}) we can write down an effective model of the single-pair neutral excitations with one quasiparticle with guiding center coordinates $R_0^a$ and two quasiholes with $R_1^a,R_2^a$ respectively. The Hamiltonian is given by
\begin{eqnarray}\label{3bdy}
\mathcal H_{phh}=&&\int \frac{d^2ql_B^2}{2\pi}
U(|q|)\left(e^{iq_a\left(R_0^a-R_1^a\right)}+e^{iq_a\left(R_0^a-R_2^a\right)}\right )\nonumber\\
&&+\int \frac{d^2ql_B^2}{2\pi} V(|q|)e^{iq_a\left(R_1^a-R_2^a\right)} .
\end{eqnarray}

\begin{figure}
\includegraphics[width=7.5cm]{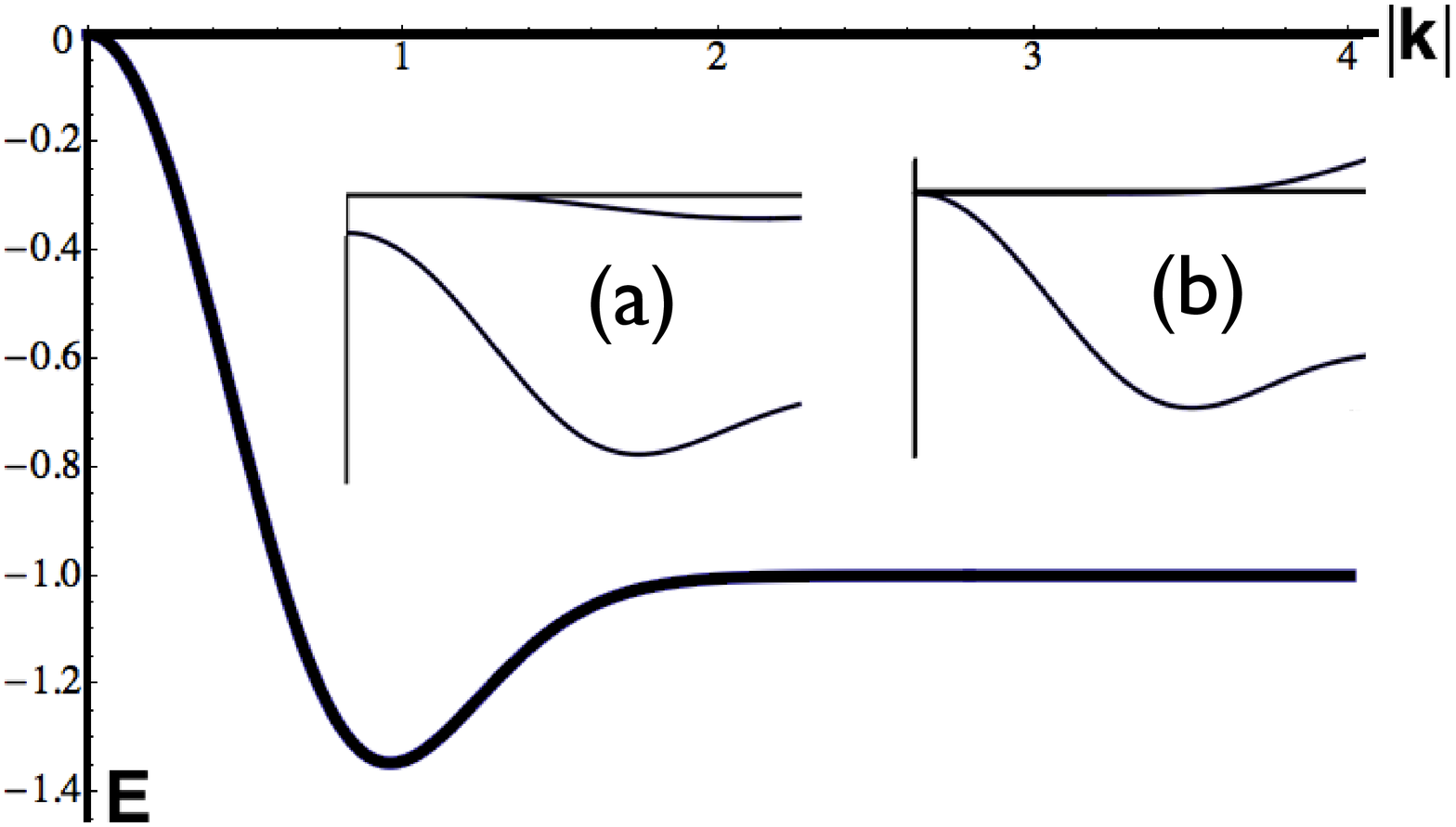}
\caption{Spectrum of  the model (\ref{3bdy}) with $E_0$ = $\infty$, $E_1$ = $-1$, $E_{n>1}$ = $V_n=0$, exhibiting a ``roton-like'' feature
at $|k|$ = $Kl_B^*$ = 0.96.
Inset (a):  $E_0$ = $\infty$, $E_1$ = $-1$, $E_2$ = $-0.1$, $E_{n>2}$ = $V_n$ =0;  inset (b): $E_0$ = $\infty$, $E_1$ = $-1$, $V_0$ = 0.1, $E_{n>1}$ = $V_{n>0}$ = 0.
}
\label{mplot}
\end{figure}

The electric dipole moment carried by the neutral excitation is $e^*R^a$, where  $R^a$ = $2R_0^a-\left(R_1^a+R_2^a\right)$, and  $[R^a,R^b]=0,[R^a,\mathcal H_{phh}]=0$.    
The residual dynamical degree of freedom is $R_{12}^a$ = $R_1^a-R_2^a$, where $R_{12}^x + iR_{12}^y$ = $2l_B^*a$, with $[a,a^{\dagger}]$ = 1.
After some algebra, (\ref{3bdy}) can be written as an effective one-body Hamiltonian
\begin{eqnarray}\label{eff2bdy}
\mathcal H_{phh}(\bm R)=\sum_n V_n|\psi^0_n\rangle\langle\psi^0_n|+\sum_{\pm}E_n|\psi^{\pm}_n\rangle\langle\psi^{\pm}_n|,  \qquad
\end{eqnarray}
where  $a|\psi_0^0\rangle$ = 0, $a^{\dagger}|\psi_n^0\rangle$ = $\surd((n+1)|\psi^0_{n+1}\rangle$, and
$|\psi_n^{\pm}\rangle$ = $\exp \pm (ka^{\dagger} - k^*a) |\psi^0_n\rangle$, with complex $k$ = $(R^x + iR^y)/l_B^*$.  Note that
the momentum $\hbar \bm K$ carried by the neutral mode is proportional to its dipole moment, with a magnitude  $K$ given by
$Kl_B^*$ = $|k|$. 
As in Ref. \cite{haldanesphere}, our formalism requires 
the state to be symmetric in the quasihole coordinates (even
under $a \rightarrow -a$), so
eigenstates of (\ref{eff2bdy}) have the form
\begin{eqnarray}\label{eigenstate}
|\psi(k)\rangle &=&\sum_{n=0}^\infty A_n|\psi^0_{2n}\rangle+B_n\left(|\psi^{+}_{2n}\rangle+|\psi^{-}_{2n}\rangle\right)
\nonumber \\
&&+C_n\left(|\psi^{+}_{2n+1}\rangle-|\psi^{-}_{2n+1}\rangle\right) .
\end{eqnarray}
In general, the model (\ref{eff2bdy}) is hard to solve analytically. It is instructive to look at the simplest case where $E_0$ = $\infty$, $E_1$ = $-1$, $E_{n>1}$ = $V_m=0$. For the pseudopotential Hamiltonian where the Laughlin state is the exact ground state, all quasihole states have zero energy; thus quasiholes do not interact with each other, at least when they are far away from quasiparticles. This, together with the quasi-degenerate energies of the  $n > 1$ quasielectrons in Fig.(\ref{qe}), suggests that the simplest case should capture most of the physics of the realistic systems. A straightforward computation shows there is only one non-zero-energy band, with an energy dispersion (relative to the sum of the energies of an isolated   $2e/3$ quasiparticle and two isolated quasiholes) 
\begin{eqnarray}\label{magband}
E(k)=S_k\left(1 -4|k|^2/(1+S_k)\right)-1 ,
\end{eqnarray}
where $S_k$ $\equiv$ $|\langle \psi^+_0|\psi^-_0\rangle |$ = $\exp -2|k|^2$. This band, shown in Fig.\ref{mplot}, qualitatively resembles the familiar magneto-roton mode\cite{gmp}, although its  minimum at
$|k|$ $\approx$ 0.96 predicts  a too-small value  $Kl_B$ = $0.96 \surd(e^*/e)$ = 0.55 (instead of $Kl_B$ $\approx$ 2) for the  $\nu$ = 1/3 roton minimum, and has a  shape that seems too pronounced as compared to Fig.\ref{nm}(a).  This discrepancy could be mainly because at the roton minimum, the separation between the quasiparticle and the quasihole is only on the order of the magnetic length, and these localized charged objects with finite sizes are less well defined. There is a non-dispersing continuum, qualitatively agreeing with the continuum of neutral excitations in Fig.\ref{nm}(a). 
When further pseudopotentials $E_{n>1}$ and $V_n$ are included, additional bands split off from the continuum: see insets in Fig.(\ref{mplot}).
At $k=0$, for each integer $m  > 0$, there is a band with energy $V_{2m}+2E_{2m}$; in general the energies of different bands split in the long wavelength limit.
While the simplest form of the ``toy model'' (\ref{3bdy}) does not quantitatively fit the collective-mode spectrum,  it  appears  to describe some of the essential physics of the neutral excitations deriving from the composite structure
of the quasielectron.

In conclusion, we have exposed an internal structure of  the  charge-$e^*$``quasielectron''  of the  Laughlin state, revealing it as a composite of a charge $2e^*$ quasiparticle and a charge $-e^*$ quasihole.  We identified the internal orbital angular-momentum of the composite with the ``$\Lambda$-level'' index in Jain's composite-fermion construction\cite{jainbook}, and constructed a simple model (\ref{3bdy})  capturing some of its features.  This structure may be relevant to the more recent  experimental results reported in Ref. \cite{west}.

\begin{acknowledgements}
This work was supported  by DOE grant DE-SC0002140 and the W. M.  Keck Foundation.   B. Y.  also acknowledges support by a  National Science Scholarship  from the Agency for Science, Technology and Research (Singapore). 
\end{acknowledgements}


\begin{thebibliography}{99}

\bibitem{laughlin}
R. B. Laughlin, Phys. Rev. Lett. {\bf 50}, 1395 (1983).


\bibitem{jain1}
G. S. Jeon and J. K. Jain, Phys. Rev. B. {\bf 68}, 165346 (2003).

\bibitem{jainbook}
J. K. Jain, ``\textit{Composite Fermions}'', Cambridge University Press, Cambridge (2007), p. 130.
 
\bibitem{hansson}
T. H. Hansson, C. C. Chang, J. K. Jain and S. Viefers, Phys. Rev. B. {\bf 76}, 075347 (2007).



\bibitem{rodriguez}
I. D. Rodriguez, A. Sterdyniak, M. Hermanns, J. K. Slingerland and N. Regnault, Phys. Rev. B. {\bf 85}, 035128 (2012).

\bibitem{bernevig}
B. A. Bernevig and F. D. M. Haldane, Phys. Rev. Lett. {\bf 102}, 066802 (2009).

\bibitem{haldanesphere} F. D. M. Haldane, Phys.Rev. Lett. {\bf  51}, 605   (1983). 

\bibitem{jain2} D. Majumder, S. S. Mandal and J. K. Jain, Nat. Phys. {\bf 5}, 403 (2009).

\bibitem{pinczuk}
C. F. Hirjibehedin, I. Dujovne, A. Pinczuk, B. S. Dennis, L. N. Pfeiffer and K. W. West, Phys. Rev. Lett. {\bf 95}, 066803 (2005).


\bibitem{regnault}    B. A. Bernevig and N. Regnault, Phys. Rev. Lett. {\bf 103}, 206801 (2009).

\bibitem{bh}
B. A. Bernevig and F. D. M. Haldane, Phys. Rev. Lett. {\bf 100}, 246802 (2008).

\bibitem{yangbo}
B. Yang, Z.-X. Hu, Z. Papic and F. D. M. Haldane, Phys. Rev. Lett. {\bf 108}, 256807 (2012).


\bibitem{gmp}
S. M. Girvin, A. H. MacDonald and P. M. Platzman, Phys. Rev. Lett. {\bf 54}, 581 (1985); Phys. Rev. B {\bf 33}, 2481 (1986).



\bibitem{west}
T. D. Rhone \textit{et al.}, Phys. Rev. Lett. {\bf 106}, 096803 (2011).





\end{thebibliography}
\end{document}